\documentclass[runningheads]{llncs}
\usepackage{graphicx}
\usepackage{amsmath}
\usepackage{amsfonts}
\usepackage{amssymb}
\usepackage{wrapfig}
\usepackage{hyperref}
\begin{document}
\title{Statistical Shape Modeling of Biventricular Anatomy with Shared Boundaries}
%
\titlerunning{SSM with Shared Boundaries}
%
\author{Krithika Iyer\inst{1,2}\and Alan Morris\inst{2} \and Brian Zenger\inst{2,4} \and
Karthik Karanth\inst{1,2} \and Benjamin A Orkild\inst{2,3} \and Oleksandre Korshak\inst{1,2} \and Shireen Elhabian\inst{1,2}}
%
\authorrunning{K. Iyer et al.}
%
\institute{University of Utah, School of Computing, Salt Lake City, UT, USA \\
\email{krithika.iyer@utah.edu} \and
University of Utah, Scientific Computing and Imaging Institute, Salt Lake City, UT, USA\\
\email{\{amorris, karthik, oleks, shireen\}@sci.utah.edu} \and 
University of Utah, Department of Biomedical Engineering, Salt Lake City, UT, USA\\
\email{ben.orkild@utah.edu } \and
University of Utah School of Medicine, Salt Lake City, UT, USA\\
\email{brian.zenger@hsc.utah.edu}}
\maketitle
\let\thefootnote\relax\footnotetext{Provisionally accepted for Statistical Atlases and Computational Modelling of the Heart Workshop at MICCAI 2022}
\begin{abstract}
Statistical shape modeling (SSM) is a valuable and powerful tool to generate a detailed representation of complex anatomy that enables quantitative analysis and the comparison of shapes and their variations. SSM applies mathematics, statistics, and computing to parse the shape into a quantitative representation (such as correspondence points or landmarks) that will help answer various questions about the anatomical variations across the population. Complex anatomical structures have many diverse parts with varying interactions or intricate architecture. For example, the heart is a four-chambered anatomy with several shared boundaries between chambers. Coordinated and efficient contraction of the chambers of the heart is necessary to adequately perfuse end organs throughout the body. Subtle shape changes within these shared boundaries of the heart can indicate potential pathological changes that lead to uncoordinated contraction and poor end-organ perfusion. Early detection and robust quantification could provide insight into ideal treatment techniques and intervention timing. 
However, existing SSM approaches fall short of explicitly modeling the statistics of shared boundaries.
In this paper, we present a general and flexible data-driven approach for building statistical shape models of multi-organ anatomies with shared boundaries that captures morphological and alignment changes of individual anatomies and their shared boundary surfaces throughout the population. 
%
We demonstrate the effectiveness of the proposed methods using a biventricular heart dataset by developing shape models that consistently parameterize the cardiac biventricular structure and the interventricular septum (shared boundary surface) across the population data.

\keywords{statistical shape modeling  \and biventricular \and cardiac MRI  \and particle-based
shape modeling  \and interventricular septum}
\end{abstract}

\section{Introduction}
Statistical shape modeling (SSM) is an important computational tool employed to discover significant shape parameters directly from medical data (such as MRI and CT scans) that can fully describe complex anatomy in the context of a population. SSM is used in biomedical research to visualize organs \cite{orkild2022all}, bones \cite{lenz2021statistical}, and tumors \cite{krol2013virtual}, aid surgical planning and guidance \cite{borghi2020population}, monitor disease progression \cite{uetani2015statistical,faber2020subregional}, and implant design \cite{goparaju2022benchmarking}. 

Traditionally, SSM approaches have focused on generating organ or disease-specific models of single-organ anatomy. However, the human body consists of complex organs and systems that are functionally, spatially, and physically connected \cite{cates2014computational,marrouche2014association,sanfilippo1990atrial}. 
%
%
Recent research in computational anatomy has shifted focus towards modeling multi-organ anatomies \cite{cerrolaza2019computational}. The motivation for modeling multiple organs stems from the need to consider joint statistical shape analysis to quantify meaningful shape variations and contextual information when studying the group differences and identifying the shape differences occurring due to a particular pathology affecting multiple interacting organs. 
Such comprehensive analysis of multiple organs and their interactions can be incredibly beneficial in diagnosing and providing timely therapeutic assistance \cite{kobatake2007future,hensel2007development,si2017point}. 
Specifically, in the case of cardiology, the interventricular septum (IVS) has been shown to change shape during various cardiomyopathies. Others have described the flattening and reversal of curvature in patients with significant right ventricular pathologies \cite{tanaka1980diastolic}. Therefore, it is crucial to model the left and right ventricle together and also the changes at the interventricular septum, or the shared boundary. 

Shapes can be represented using an implicit (deformation fields \cite{durrleman2014morphometry}, level set methods \cite{samson2000level}) or explicit (landmarks/points) representation. Explicit parameterization, such as landmarks, is one of the most popular techniques used to represent shapes because of its simplicity and ability to represent multiple objects easily \cite{cerrolaza2019computational}. Hence, in this work, we focus on point distribution models (PDM) for representation as PDMs are suitable for the statistical analysis of models with shared boundaries. To enable comparison and to obtain shape statistics in an ensemble of shapes, points of the same anatomical position must be established consistently across shape populations. These points are called \textit{correspondences}. Multiple methods for correspondence generation have been proposed, which include non-optimized landmark estimation, parametric, and non-parametric correspondence optimization. Non-optimized methods entail manually annotating the reference shape and warping these landmarks on the population data, and they employ hard surface constraints to distribute points on a shape. Parametric methods use fixed geometrical basis (e.g., spheres) \cite{styner2006framework} to parameterize objects and generate correspondences. The correspondence model obtained using manual or parametric techniques is not optimal and can prove to be incapable of handling complex anatomies. On the other hand, non-parametric methods provide a robust and general framework as they use a PDM without relying on a specific geometric basis. Methods that follow group-wise approach find the correspondence by considering the variability of entire data in the optimization process (e.g., particle-based optimization \cite{cates2017shapeworks}, Minimum Description Length - MDL \cite{davies2002learning}).
The group-wise SSM approaches have been extended to model multi-organ anatomies. These approaches either parameterize each object separately, sacrificing anatomical integrity \cite{cerrolaza2019computational}, or minimize the combined cost function to generate correspondences assuming a global statistical model \cite{cates2008particle,durrleman2014morphometry}.

To the best of our knowledge, none of the existing SSM methods have explicitly incorporated nuanced interactions such as shared surfaces between multiple anatomies that can reveal key features that might not be observable on the individual anatomies when modeled independently. To address this issue, we propose a mesh grooming pipeline for extracting shared boundary surfaces and a correspondence-based optimization scheme to parameterize multi-organ anatomies and their shared surfaces consistently. We demonstrate the entire pipeline with a cardiac biventricular dataset, where we model the right ventricle (RV), left ventricle wall (LVW), and the interventricular septum (IVS). We use the group-wise non-parametric particle-based optimization method proposed by Cates et al., \cite{cates2007shape,cates2017shapeworks} to generate PDM and modify the framework to support multi-organ anatomies with shared boundaries.  
%
%


\section{Methods}

Constructing a statistical shape model for multi-organ anatomies with shared boundaries requires consistent point distribution on the shared boundary across the multi-organ anatomies and explicitly modeling the statistics of both the contour and the interior of the shared boundary. 
To fulfill these requirements, we first need tools to detect and extract shared boundaries and their edges (i.e., contour information) from two adjoining anatomies. The steps and methods for the proposed general pipeline for shared boundary extraction are explained in Section \ref{shared_boundary_tool}. 
Second, we need to fit a PDM that includes joint statistics of the multi-organ anatomies,
shared boundary interior and contour. Herein we leverage the particle-based shape modeling (PSM) approach \cite{cates2007shape,cates2017shapeworks} for automatically constructing PDMs by optimizing point (or particle) distributions over a cohort of shapes using an entropy-based optimization method.
The PSM method uses a system of interacting particles with mutually repelling forces that learn the most compact statistical descriptors of the anatomy \cite{cates2017shapeworks}. For consistent parameterization on the shared boundary, the surface sampling objective of the PSM method has to be modified to accommodate the interaction between the anatomies and the shared surface. A brief overview of the PSM entropy optimization method for single anatomy is provided in Section \ref{entropy_based_optimization} and the proposed surface cost function modifications for multi-organ anatomies with shared boundary surfaces is provided in Section \ref{entropy_based_shared_boundary}.

\subsection{Background: Particle-based Shape Modeling}\label{entropy_based_optimization}
Consider a cohort of shapes \(\mathcal{S} = \{\mathbf{z}_1, \mathbf{z}_2, ..., \mathbf{z}_N \}\) of \(N\) surfaces, each with its set of \(M\) corresponding particles \(\mathbf{z}_n = \left[\mathbf{x}_1, \mathbf{x}_2, ..., \mathbf{x}_M \right] \in \mathbb{R}^{dM} \) where each particle  \(\mathbf{x}_m \in \mathbb{R}^d\) lives in \(d-\)dimensional Cartesian (i.e., configuration) space. The ordering of the particles implies correspondence among shapes. Each of the correspondence particles is constrained to lie on the shape's surface. Collectively, the set of \(M\) particles is known as the \textit{configuration} and the space of all possible configurations is known as the \textit{configuration space}. The particle positions are samples (i.e., realizations) of a random variable \(\mathbf{X} \in \mathbb{R}^d\) in the configuration space with an associated probability distribution function (PDF) \(p(\mathbf{X=x})\). Each configuration of \(M\) particles can be mapped to a single \(dM-\)dimensional \textit{shape space} by concatenating the correspondence coordinate positions into a single vector \(\mathbf{z}_n\) which is modeled as an instance of random variable \(\mathbf{Z}\) in the shape space with PDF \(p(\mathbf{Z=z})\) assuming shapes are Gaussian distributed in the shape space, i.e., \(\mathbf{Z} \sim \mathcal{N}(\boldsymbol{\mu},\boldsymbol{\Sigma})\). The optimization to establish correspondence minimizes the energy function 
\begin{equation}\label{equation_1}
    Q = H(\mathbf{Z}) - \sum_{k=1}^N H(\mathbf{X}_k)
\end{equation}
where \(H\) is an estimation of differential entropy. The differential entropy of \(p(\mathbf{X})\) is given as
\begin{equation}\label{equation_2}
    H(\mathbf{X}) = -\int_S p(\mathbf{X}) \log p(\mathbf{X}) dx = - E\{\log p(\mathbf{X})\} \approx - \frac{1}{M} \sum_{i=1}^M \log p(\mathbf{x}_i)
\end{equation}

Minimization of the first term in \(Q\) from Eq (\ref{equation_1}) produces a compact distribution of samples in shape space and encourages particles to be in correspondence across shapes. The second term seeks uniformly-distributed correspondence positions on the shape surfaces for accurate shape representation \cite{cates2007shape,cates2017shapeworks}. Further details regarding the optimization and gradient updates can be found in \cite{cates2007shape,cates2017shapeworks}. 

\subsection{Shared Boundary Extraction}\label{shared_boundary_tool}
\begin{wrapfigure}[22]{R}{0.5\textwidth}
    \centering
    \includegraphics[scale=0.4]{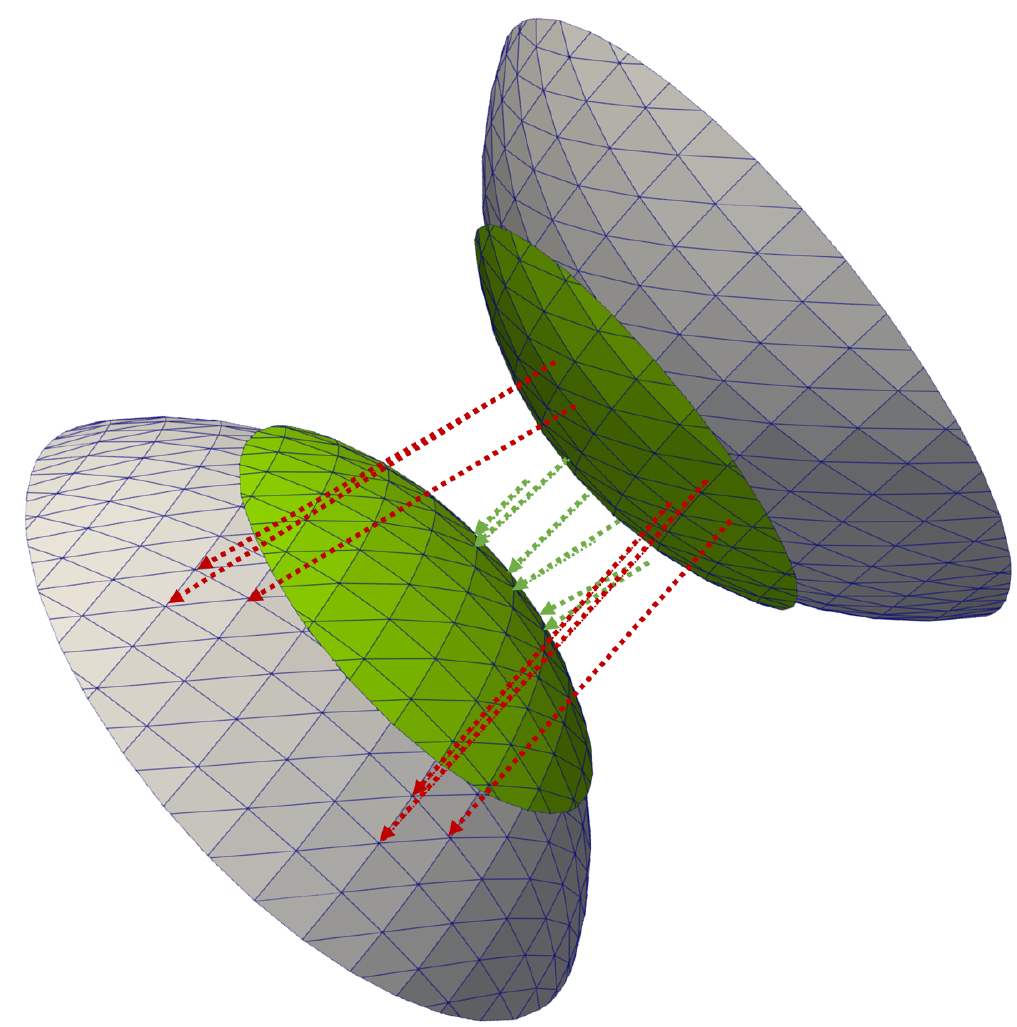}
    \caption{Extracting shared boundary between two meshes. The regions in green have Euclidean distance that fall within the threshold and are extracted as shared boundary as per step 2. The green arrows show the distances within the threshold and the red arrows show distances greater than the threshold. The contour is extracted from the green region as per step 4. Note: the meshes are farther apart and the threshold is larger for visualization purposes.}
    \label{fig:extracting_shared_boundary}
\end{wrapfigure}

To demonstrate the shared boundary extraction pipeline, consider two adjoining organs \(A\) and \(B\), with a shared boundary. The steps for shared boundary extraction entails:
\begin{enumerate}
\item{\textbf{Isotropic Explicit Re-meshing:}} This generates a new mesh triangulation that conforms to the original data, but contains more uniformly sized triangles. This also has the benefit of ensuring equivalent average edge lengths across the two shapes, which is useful in ensuing steps \cite{surazhsky2003isotropic}.
\item{\textbf{Extracting Shared Boundary:}} In this step, we ingest the two original shapes and output three new shapes, two of which correspond to the original shapes and one for the shared boundary. Let us designate the original meshes as \(A_o\) and \(B_o\) (Figure \ref{fig:extracting_shared_boundary}.a and \ref{fig:extracting_shared_boundary}.b) then:
\begin{enumerate}
    \item Find all the triangles in \(A_o\) that are close to \(B_o\) and construct a mesh with these triangles called \(A_s\). A triangle with vertices \(v_0,v_1\) and \(v_2\) is considered close to a mesh if the shortest Euclidean distance to the mesh for all the three vertices is below a small threshold. The threshold has to experimentally tuned in order to ensure the extracted shared surfaces are clinically relevant. We similarly find all the triangles in \(B_o\) that are close to \(A_o\) and designate this mesh as \(B_s\)
    \item Find the remainder of the mesh in \(A_o\) after removing the triangles in \(A_s\) and designate this as \(A_r\). Similarly, we designate the remainder of the mesh in \(B_o\) after removing the triangles in \(B_s\) as \(B_r\).
    \item Arbitrary designate \(B_s\) as the shared surface \(M\)
    \item Move all the points on the boundary loop of \(A_r\) to the boundary loop of \(M\) and return three new shapes \(A_r\), \(M\), and \(B_r\) (Figure \ref{fig:extracting_shared_boundary}.c).
\end{enumerate}

\item{\textbf{Laplacian Smoothing:}} At this point, the resulting triangulation typically contains jagged edges. We apply Laplacian smoothing to correct for this \cite{field1988laplacian}.

\item{\textbf{Extract Contour:}} The boundary loop of the shared surface \( M\) is computed using LibIGL \href{https://libigl.github.io/libigl-python-bindings/igl_docs/#boundary_loop}{\textit{boundary\_loop}} tool \cite{libigl} and designate this contour as \(C\) (Figure \ref{fig:extracting_shared_boundary}.d).
\end{enumerate}
The input consisting of two adjoining organs with a shared surface has been converted into input with four separate parts, the organs \(A\) and \(B\), the shared surface, and the contour using the pipeline (Figure \ref{fig:extracting_shared_boundary}.d). 

\begin{figure}
    \centering
    \includegraphics[scale=0.7]{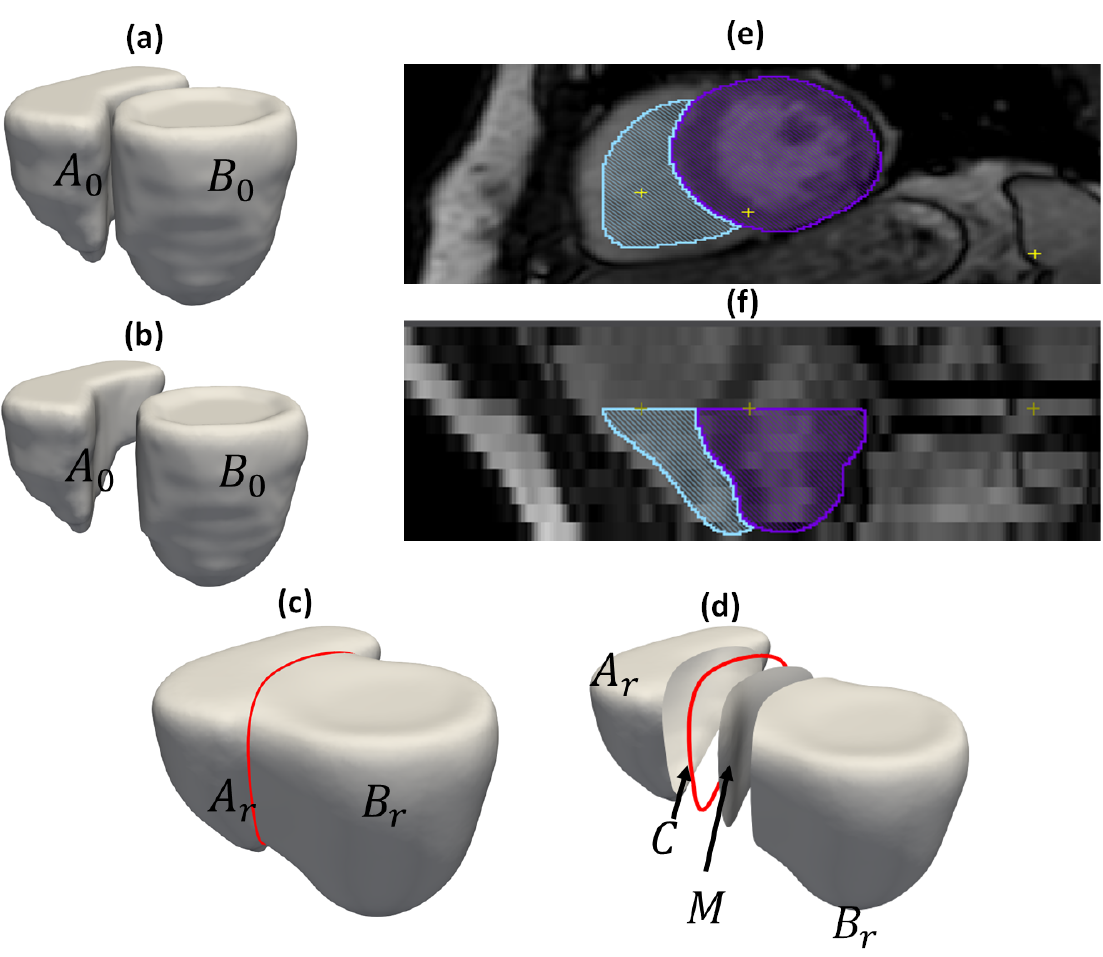}
    \caption{An example of output obtained after shared boundary extraction. Meshes representing (a) RV and LVW show that they have a shared boundary surface, and (b) RV and LVW meshes are pried apart. The meshes and contour obtained after shared boundary extraction (c) RV, LVW, shared surface and contour (d) all outputs pried apart for visualization. The red color indicates the contour. The image shows the endocardial segmentation for the RV (blue) and epicardial segmentation for the LV (violet)  at end diastole in the (e) axial view and (f) coronal view}
    \label{fig:mesh_grooming_pipeline}
\end{figure}
\subsection{Particle-based Shape Modeling with Shared Boundaries}\label{entropy_based_shared_boundary}
In section \ref{shared_boundary_tool}, two adjoining organs with shared surfaces were separated into four separate parts. Now a shape model has to be built that can faithfully capture the joint statistics of all the organs while representing the individual parts consistently. The first requirement for a shared boundary shape model is that particle-based optimization should handle multi-organ anatomies. The optimization set up in Eq (\ref{equation_1}) was extended for multiple organs by treating all the organs as a single structure \cite{cates2008particle}.
From the original formulation, it is important to note that \(p(\mathbf{x}_i)\) in Eq \ref{equation_2} was estimated from the particle position using non-parametric kernel density estimation method \cite{cates2007shape,cates2017shapeworks}.
This results in a set of points on the surface that repel each other with Gaussian-weighted forces.
Therefore, for multiple organ anatomy, if one organ has distinct identities, the spatial interactions between particles on different organs are decoupled, and particles are constrained to lie on a single organ (surface). The covariance \(\boldsymbol{\Sigma}\) includes all particle positions across the multiple organs so that optimization takes place in the joint, multi-organ shape space and the shape statistics remains coupled \cite{cates2008particle}. 
For \(D\) organs in an anatomy, the cost function is
\begin{equation}\label{equation_3}
     Q = H(\mathbf{Z}) - \sum_{j=1}^D {\left[\sum_{k=1}^N H(\mathbf{X}^j_k)\right]}
\end{equation}
where \(\mathbf{X}^j_k\) represents the \(k^{th}\) particle on \(j^{th}\) organ.

Second, from Eq (\ref{equation_3}), it can be noted that the second term, which represents the sampling objective, is summed over all the shape samples, such that the sampling is restricted to the particles contained within the individual organs. As a result, when two organs have a shared boundary and sampling is done independently, there is no mechanism to ensure that the particles do not clutter around the edges of the organs. Hence, the sampling objective needs to be modified such that the particles on the shared boundary contour repel the particles of other mesh objects. This will result in a buffer distance between particles of the multiple organs leading to a uniform correspondence model.
%
The proposed objective function is:
\begin{equation}\label{equaton_5}
     Q = H(\mathbf{Z}) -  {\left[\sum_{j \in (A_r, M, B_r)}\sum_{k=1}^N H\binom{\mathbf{X}^j_k}{\mathbf{X}^C_k} + \sum_{k=1}^N H(\mathbf{X}^C_k)\right]}
\end{equation}
where \(\mathbf{X}^c\) is the matrix of particle positions located in the contour. Effectively, this means that all
the particles on the \(A_r\), \(M\) and \(B_r\) are repelled by the contour \(C\) particles. We do not change the sampling objective for the contour. This is because large gradients from the meshes could cause the particles on the contour
to swap places. Since there is only one degree of freedom on a contour, it is almost impossible to recover from this situation.


\section{Experiments and Results}
\quad \textbf{Dataset:} We evaluate our method on a cardiac biventricular dataset based on how the resulting correspondence model captures variability in shape for cardiovascular clinic patients and healthy volunteer groups. The dataset consists of MRIs of 6 healthy volunteers and 23 patients treated at a cardiovascular clinic. In the patient group, tricuspid regurgitation was secondary to pulmonary hypertension in one patient; congestive heart failure (CHF) in 10 patients; and other causes (atrial fibrillation, pacemaker lead injury, pacemaker implantation, congenital heart disease) in 12 patients. The healthy volunteers had no diagnosis of cardiac disease and no cardiovascular risk factors. 

Initially, the RV and LVW segmentation images were generated by converting end-diastole CINE MRI to volume stack. From each CINE short axis time stack, an image of the heart at end diastole was extracted to create a volume image stack. Image extraction was performed using a custom MATLAB image processing code. The volume stacks were then segmented using the open-source Seg3D software (SCI Institute, University of Utah, SLC UT). The segmentations were then isotropically resampled and converted to meshes using the open software \href{http://sciinstitute.github.io/ShapeWorks/}{ShapeWorks}. In order to align the shapes, the meshes were centered and rigidly aligned to a representative reference sample selected from the population. The rigid alignment was done by calculating the transformations only using the RV meshes of the population due to their complex shapes. These transformations were then applied to the RV and the LVW meshes. The average edge length of the right ventricle meshes was \(0.8224 \pm 0.3987\), left ventricle wall meshes was \(0.9438 \pm 0.3399\), the IVS meshes \(0.5196 \pm 0.4047\), and the contours \(21.469 \pm 26.205\).

\textbf{Shape Model Construction:} We used \href{http://sciinstitute.github.io/ShapeWorks/}{ShapeWorks}, an open-source software that implements the particle-based entropy optimization \cite{cates2007shape,cates2017shapeworks} described in section \ref{entropy_based_optimization}. We modified the optimization with the proposed cost function (equation \ref{equaton_5}) to support multi-organ anatomies with shared boundaries. First, the shared boundary surface and contour were extracted for building a shape model using the tool described in section \ref{shared_boundary_tool}. Figure.\ref{fig:mesh_grooming_pipeline} shows an example output for one sample. Then, a shape model was built using 512 particles for the RV and LVW, and 64 particles were used for the IVS surface and contour. From this PDM, mean shapes and differences were computed.

\begin{figure}[!ht]
    \centering
    \includegraphics[scale=0.45]{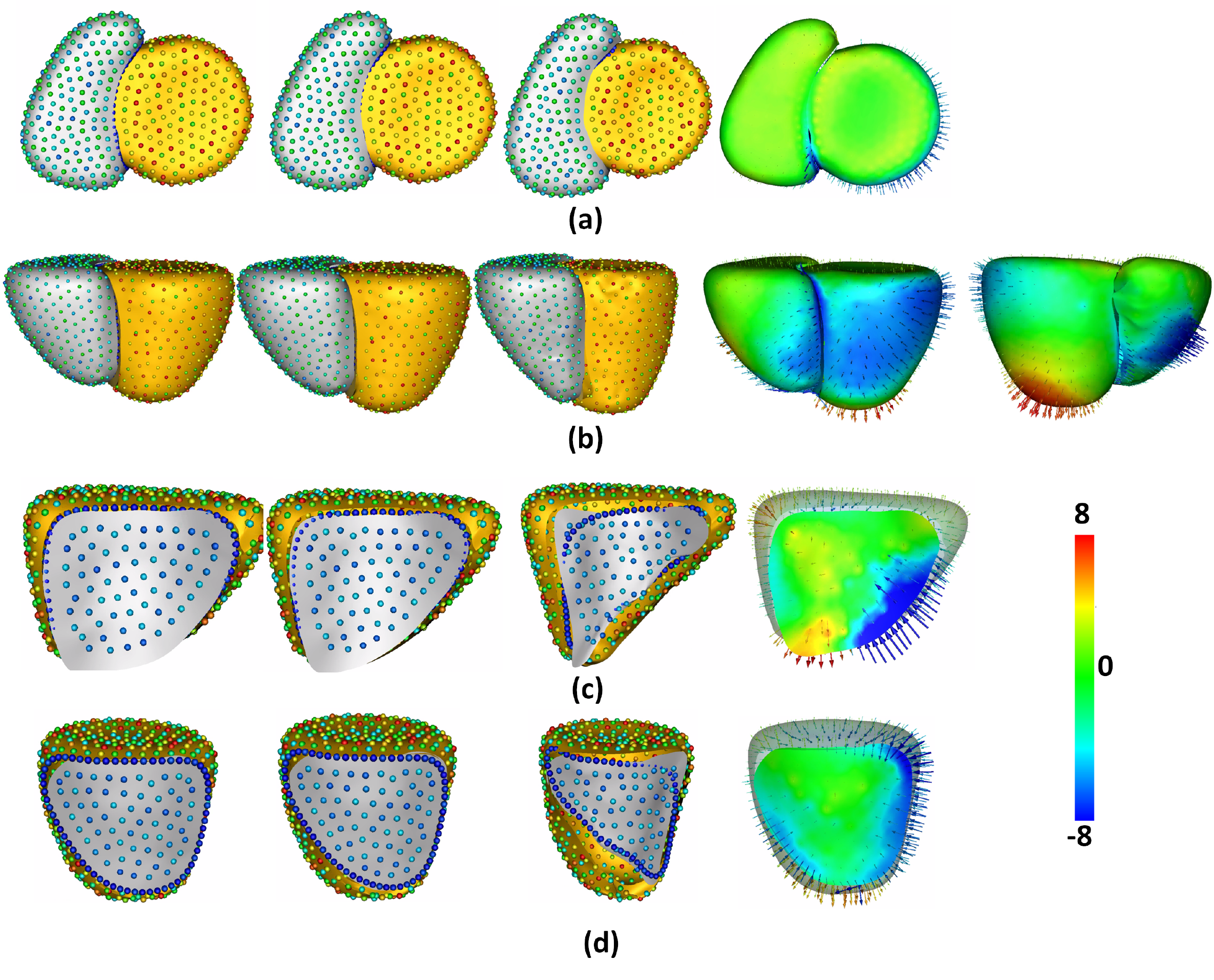}
    \caption{Columns 1, 2, and 3 show the mean shape of the patient group, overall mean, and mean shape of the control group, respectively. Columns 4 and 5 show the difference between the group-mean shapes (two views). The arrows indicate the direction of group differences, and the color represents the magnitude of the group difference. The PDM for bivenctricle data (a) top-view and (b) front view. The same shapes after (c) excluding the left ventricular wall and (d) excluding the right ventricle for visualizing the IVS.}
    \label{figure_group_differences}
\end{figure}

\begin{figure}
    \centering
    \includegraphics[scale=0.3]{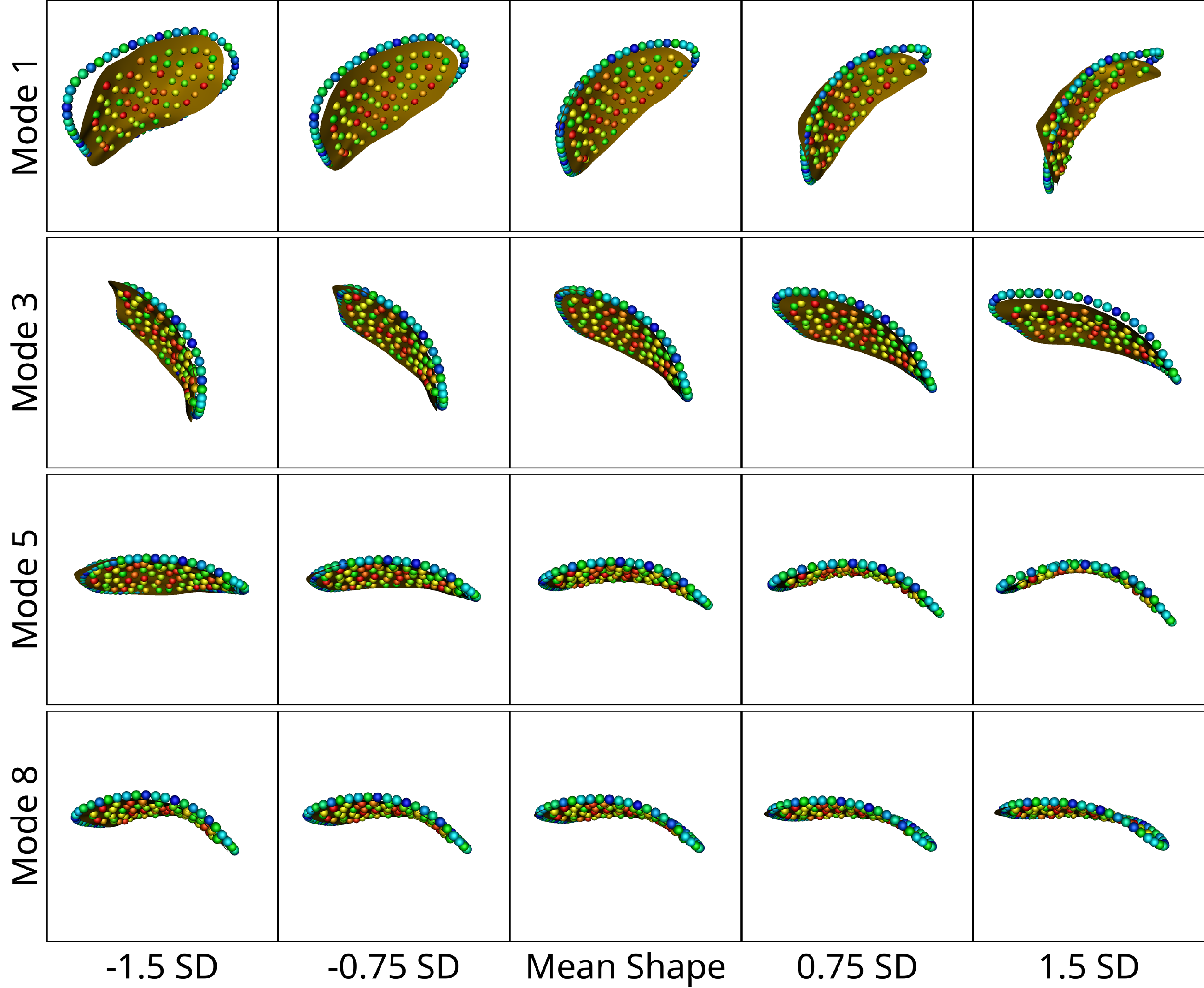}
    \caption{Modes of variation of the IVS shared surface and contour that capture the change in curvature. The model was obtained using all four parts of the anatomy - RV, LVW, IVS shared surface, and contour. The RV and LVW are excluded only for visualization.}
    \label{fig:ivs_modes}
\end{figure}

\textbf{Discussion:} The shape model was used to identify group-level shape differences of the RV, LVW, and IVS. Figure.\ref{figure_group_differences} shows the mean shape of each group and the color-coded group differences. There is a marked difference in the curvature of IVS of the healthy group as compared to the patient group. The curvature of the IVS was also captured in the modes of variation of the shape model. In Figure.\ref{fig:ivs_modes}, we visualize the modes of variation obtained using principal component analysis of the IVS shared surface and contour obtained after building the shape model with all the four parts - RV, LVW, IVS shared surface, and contour. The RV and LVW are excluded only for visualization. Since the curvature is not a linear feature, a single PCA mode is not enough to capture it. We show the top four PCA modes that capture the curvature in various directions. 

\begin{figure}
    \centering
    \includegraphics[scale=0.4]{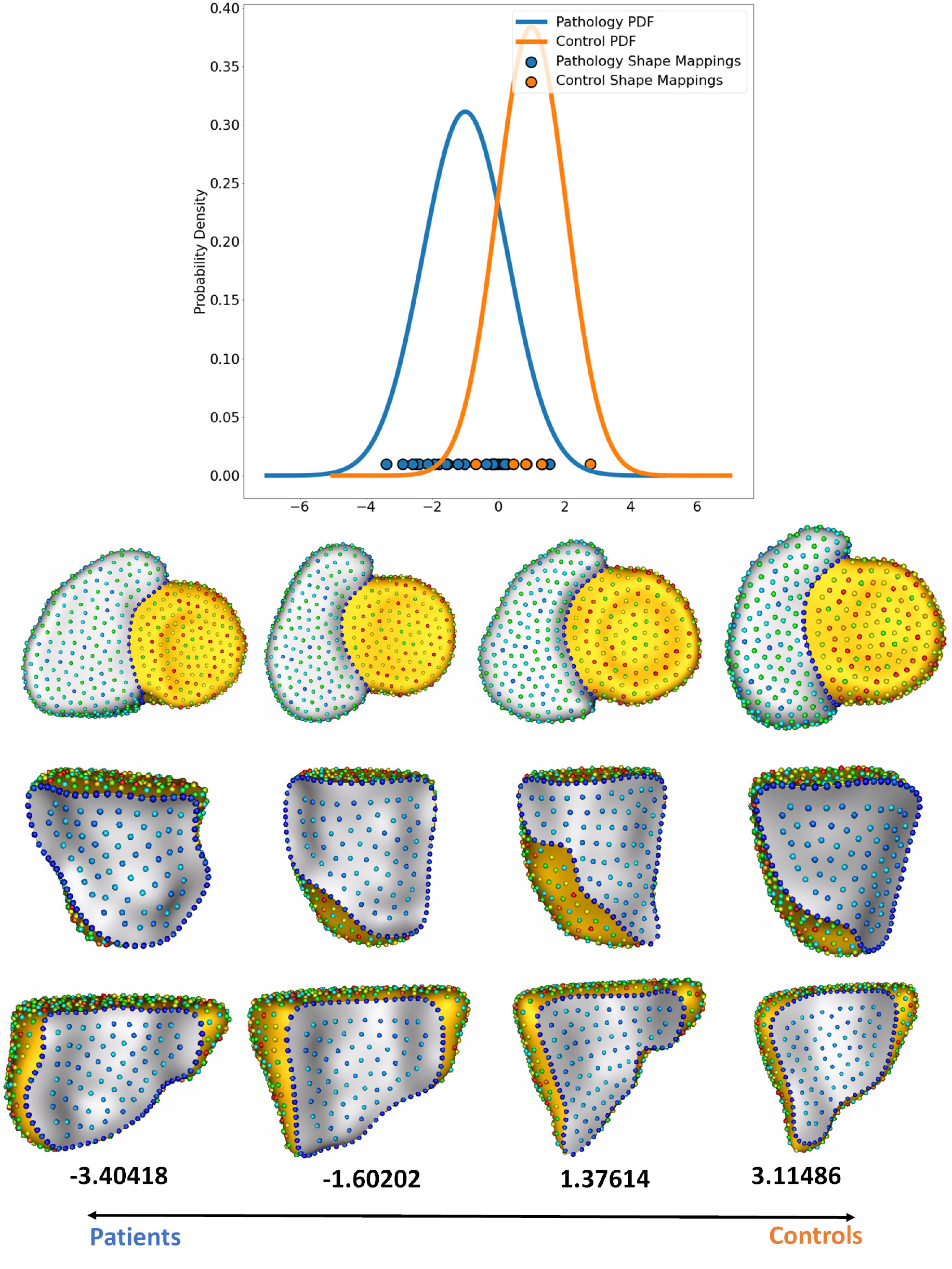}
    \caption{Shape mapping to linear discrimination of variation between population means for the groups of patients and controls. The first row represents the biventricle shared boundary shapes. The second row represents the same biventricle shared boundary shapes after excluding the left ventricular wall, and the third row after excluding the right ventricle for visualizing the IVS. The number below each shape denotes the “shape-based score” of each anatomy derived from the shape population.}
    \label{figure_lda}
\end{figure}

In order to study statistically significant geometric differences, we performed linear discrimination of variation. The particle-wise mean shapes of both groups were compared, and a difference vector was generated. The group means for the cardiac patients is set as -1, and controls are set as 1. Each shape is mapped to a single scalar value (or a "shape-based-score") that places subject-specific anatomy on a group-based shape difference statistically derived from the shape population. Figure.\ref{figure_lda} shows the mapping for all shapes of the two groups. Selected shapes correspond to individual points on the graph. The shapes at the extreme ends of the mapping also confirm that the shape model appropriately identified the curvature of IVS as a significant geometrical difference between the two groups. The shape in Figure.\ref{figure_lda} also shows free wall bulging and narrowing of the base of RV for cardiac patients. For modes of variation of the shape model, see Appendix.\ref{figure_modes_of_variation}.

\begin{figure}
    \centering
    \includegraphics[scale=0.3]{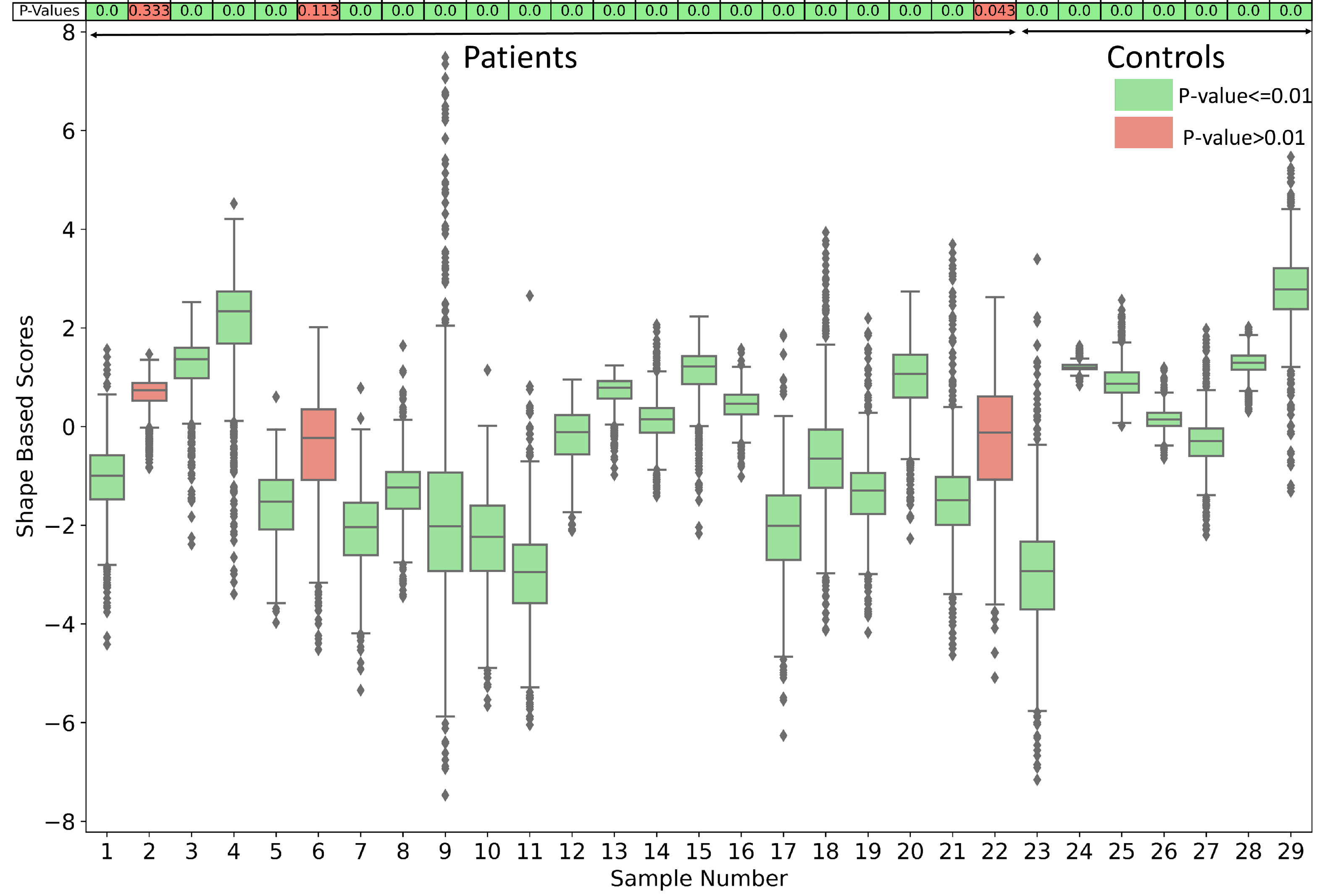}
    \caption{Statistical test for testing the effect of dataset imbalance on the shape-based scores. Box and whisker plot showing the distribution of shape-based scores for each sample obtained using a subset of the patient group, and all the control samples repeated 1000 times. Each box is color-coded based on the p-values: green - samples with p-values\(\leq 0.01\) and red -samples with p-value\(>0.01\).}
    \label{fig:t_test}
\end{figure}

Since the number of samples in the patient group and control group are not the same, we performed hypothesis testing to identify if the shape-based score assigned to each sample is statistically significant and agnostic to the data imbalance. We generated the shape-based scores for each sample using the statistics of 6 randomly selected samples from the patient group and all six control group samples and repeated the experiment 1000 times. The shape-based scores from the experiment were then compared to those generated using the complete dataset. We use t-test to test for the null hypothesis that the expected value (mean) of a sample of independent observations from the 1000 trials is equal to the given population mean, i.e., the scores generated using the complete dataset. Figure.\ref{fig:t_test} shows the box-and-whisker plot of the distribution of scores of each sample obtained from the experiment, and the color indicates the p-values. We select the alpha value to be 0.01. Hence, if the p-values are smaller than 0.01, the null hypothesis holds (shown in green), and if the p-value is greater than 0.01, we can reject the null hypothesis and assume that the scores are affected by the imbalance (shown in red). It can be seen from Figure.\ref{fig:t_test} that the imbalance does not affect the shape-based scores for the majority of samples.

These results confirm what has been observed in the cardiology literature: a decrease in interventricular septal curvature during prolonged right ventricular dysfunction. A healthy heart has a significant pressure gradient between the right and left ventricles. However, in many cardiac diseases, the pressure gradient dissipates because right ventricular pressure increases. As the pressure increases, a distortion occurs at the interventricular septum, and the original septal curvature matching the left ventricular becomes flattened. This signifies the structural remodeling that occurs with severe right-sided cardiac pathologies.

Despite these observations being made previously, the clinical utility of septal curvature has been minimal because of inadequate tools for precise and accurate measurements. Structural remodeling initially occurs to compensate for acute changes in cardiac physiology. As acute changes become chronic, the cardiac structural adaptations become permanent and cause long-term detrimental effects. Initially, patients do not feel significant symptoms because of cardiac tissue's excellent adaptability and plasticity. However, structural changes, like the ones noted above, are often already present and easily detectable. Therefore, shape analysis of interventricular septal curvature changes could be used as an early prognosticator of cardiac dysfunction prior to patients reporting significant symptoms. Notably, shape analysis can be performed using non-invasive imaging and does not require cardiac catheterization, a routine, invasive diagnostic procedure typically used for detecting cardiac dysfunction. Thus, the proposed shared-boundary SSM generation technique can potentially improve patient outcomes with early diagnosis using non-invasive imaging procedures.

\section{Conclusion}
Our method provides a novel way of extracting and generating shape models of multi-organ anatomies with shared boundary surfaces. We showed our method provides a consistent and robust representation of the shared boundary without compromising the integrity of the multiple-organ PDM. We applied our method to a cardiac biventricular dataset and showed unique shape changes of the IVS that is not captured when modeling the ventricles alone. This pipeline could pave the way for using shape analysis from non-invasive imaging for early diagnosis and prognostication of pathologies affecting multiple organs and further our understanding of interactions between any anatomical system with shared boundaries.

\section{Acknowledgement}
This work was supported by the National Institutes of Health under grant numbers NIBIB-U24EB029011, NIAMS R01AR076120, NHLBI-R01HL135568, NIBIB-R01EB016701, NIGMS-P41GM103545, NIGMS-R24GM136986 
(MacLeod), and NHLBI-F30HL149327 (Zenger). We thank the University of Utah Division of Cardiovascular Medicine and the ShapeWorks team.
\clearpage
\bibliographystyle{splncs04}
\bibliography{refs}

\begin{thebibliography}{10}
\providecommand{\url}[1]{\texttt{#1}}
\providecommand{\urlprefix}{URL }
\providecommand{\doi}[1]{https://doi.org/#1}

\bibitem{borghi2020population}
Borghi, A., Rodriguez~Florez, N., Ruggiero, F., James, G., O’Hara, J., Ong,
  J., Jeelani, O., Dunaway, D., Schievano, S.: A population-specific material
  model for sagittal craniosynostosis to predict surgical shape outcomes.
  Biomechanics and modeling in mechanobiology  \textbf{19}(4),  1319--1329
  (2020)

\bibitem{cates2014computational}
Cates, J., Bieging, E., Morris, A., Gardner, G., Akoum, N., Kholmovski, E.,
  Marrouche, N., McGann, C., MacLeod, R.S.: Computational shape models
  characterize shape change of the left atrium in atrial fibrillation. Clinical
  Medicine Insights: Cardiology  \textbf{8},  CMC--S15710 (2014)

\bibitem{cates2017shapeworks}
Cates, J., Elhabian, S., Whitaker, R.: Shapeworks: Particle-based shape
  correspondence and visualization software. In: Statistical Shape and
  Deformation Analysis, pp. 257--298. Elsevier (2017)

\bibitem{cates2008particle}
Cates, J., Fletcher, P.T., Styner, M., Hazlett, H.C., Whitaker, R.:
  Particle-based shape analysis of multi-object complexes. In: International
  Conference on Medical Image Computing and Computer-Assisted Intervention. pp.
  477--485. Springer (2008)

\bibitem{cates2007shape}
Cates, J., Fletcher, P.T., Styner, M., Shenton, M., Whitaker, R.: Shape
  modeling and analysis with entropy-based particle systems. In: Biennial
  International Conference on Information Processing in Medical Imaging. pp.
  333--345. Springer (2007)

\bibitem{cerrolaza2019computational}
Cerrolaza, J.J., Picazo, M.L., Humbert, L., Sato, Y., Rueckert, D., Ballester,
  M.{\'A}.G., Linguraru, M.G.: Computational anatomy for multi-organ analysis
  in medical imaging: A review. Medical Image Analysis  \textbf{56},  44--67
  (2019)

\bibitem{davies2002learning}
Davies, R.H.: Learning shape: optimal models for analysing natural variability.
  The University of Manchester (United Kingdom) (2002)

\bibitem{durrleman2014morphometry}
Durrleman, S., Prastawa, M., Charon, N., Korenberg, J.R., Joshi, S., Gerig, G.,
  Trouv{\'e}, A.: Morphometry of anatomical shape complexes with dense
  deformations and sparse parameters. NeuroImage  \textbf{101},  35--49 (2014)

\bibitem{faber2020subregional}
Faber, B.G., Bredbenner, T., Baird, D., Gregory, J., Saunders, F., Giuraniuc,
  C., Aspden, R., Lane, N., Orwoll, E., Tobias, J.H., et~al.: Subregional
  statistical shape modelling identifies lesser trochanter size as a possible
  risk factor for radiographic hip osteoarthritis, a cross-sectional analysis
  from the osteoporotic fractures in men study. Osteoarthritis and cartilage
  \textbf{28}(8),  1071--1078 (2020)

\bibitem{field1988laplacian}
Field, D.A.: Laplacian smoothing and delaunay triangulations. Communications in
  applied numerical methods  \textbf{4}(6),  709--712 (1988)

\bibitem{goparaju2022benchmarking}
Goparaju, A., Iyer, K., Bone, A., Hu, N., Henninger, H.B., Anderson, A.E.,
  Durrleman, S., Jacxsens, M., Morris, A., Csecs, I., et~al.: Benchmarking
  off-the-shelf statistical shape modeling tools in clinical applications.
  Medical Image Analysis  \textbf{76},  102271 (2022)

\bibitem{hensel2007development}
Hensel, J.M., M{\'e}nard, C., Chung, P.W., Milosevic, M.F., Kirilova, A.,
  Moseley, J.L., Haider, M.A., Brock, K.K.: Development of multiorgan finite
  element-based prostate deformation model enabling registration of endorectal
  coil magnetic resonance imaging for radiotherapy planning. International
  Journal of Radiation Oncology* Biology* Physics  \textbf{68}(5),  1522--1528
  (2007)

\bibitem{libigl}
Jacobson, A., Panozzo, D., et~al.: {libigl}: A simple {C++} geometry processing
  library (2018), https://libigl.github.io/

\bibitem{kobatake2007future}
Kobatake, H.: Future cad in multi-dimensional medical images:--project on
  multi-organ, multi-disease cad system--. Computerized Medical Imaging and
  Graphics  \textbf{31}(4-5),  258--266 (2007)

\bibitem{krol2013virtual}
Krol, Z., Skadlubowicz, P., Hefti, F., Krieg, A.H.: Virtual reconstruction of
  pelvic tumor defects based on a gender-specific statistical shape model.
  Computer aided surgery  \textbf{18}(5-6),  142--153 (2013)

\bibitem{lenz2021statistical}
Lenz, A.L., Kr{\"a}henb{\"u}hl, N., Peterson, A.C., Lisonbee, R.J., Hintermann,
  B., Saltzman, C.L., Barg, A., Anderson, A.E.: Statistical shape modeling of
  the talocrural joint using a hybrid multi-articulation joint approach.
  Scientific Reports  \textbf{11}(1),  1--14 (2021)

\bibitem{marrouche2014association}
Marrouche, N.F., Wilber, D., Hindricks, G., Jais, P., Akoum, N., Marchlinski,
  F., Kholmovski, E., Burgon, N., Hu, N., Mont, L., et~al.: Association of
  atrial tissue fibrosis identified by delayed enhancement mri and atrial
  fibrillation catheter ablation: the decaaf study. Jama  \textbf{311}(5),
  498--506 (2014)

\bibitem{orkild2022all}
Orkild, B.A., Zenger, B., Iyer, K., Rupp, L.C., Ibrahim, M.M., Khashani, A.G.,
  Perez, M.D., Foote, M.D., Bergquist, J.A., Morris, A.K., et~al.: All roads
  lead to rome: Diverse etiologies of tricuspid regurgitation create a
  predictable constellation of right ventricular shape changes. Frontiers in
  Physiology p.~1092 (2022)

\bibitem{samson2000level}
Samson, C., Blanc-F{\'e}raud, L., Aubert, G., Zerubia, J.: A level set model
  for image classification. International journal of computer vision
  \textbf{40}(3),  187--197 (2000)

\bibitem{sanfilippo1990atrial}
Sanfilippo, A.J., Abascal, V.M., Sheehan, M., Oertel, L.B., Harrigan, P.,
  Hughes, R.A., Weyman, A.E.: Atrial enlargement as a consequence of atrial
  fibrillation. a prospective echocardiographic study. Circulation
  \textbf{82}(3),  792--797 (1990)

\bibitem{si2017point}
Si, W., Heng, P.A., et~al.: Point-based visuo-haptic simulation of multi-organ
  for virtual surgery. Digital Medicine  \textbf{3}(1), ~18 (2017)

\bibitem{styner2006framework}
Styner, M., Oguz, I., Xu, S., Brechb{\"u}hler, C., Pantazis, D., Levitt, J.J.,
  Shenton, M.E., Gerig, G.: Framework for the statistical shape analysis of
  brain structures using spharm-pdm. The insight journal (1071), ~242 (2006)

\bibitem{surazhsky2003isotropic}
Surazhsky, V., Alliez, P., Gotsman, C.: Isotropic remeshing of surfaces: a
  local parameterization approach. Ph.D. thesis, INRIA (2003)

\bibitem{tanaka1980diastolic}
Tanaka, H., Tei, C., Nakao, S., Tahara, M., Sakurai, S., Kashima, T., Kanehisa,
  T.: Diastolic bulging of the interventricular septum toward the left
  ventricle. an echocardiographic manifestation of negative interventricular
  pressure gradient between left and right ventricles during diastole.
  Circulation  \textbf{62}(3),  558--563 (1980)

\bibitem{uetani2015statistical}
Uetani, M., Tateyama, T., Kohara, S., Tanaka, H., Han, X.H., Kanasaki, S.,
  Furukawa, A., Chen, Y.W.: Statistical shape model of the liver and its
  application to computer-aided diagnosis of liver cirrhosis. Electrical
  Engineering in Japan  \textbf{190}(4),  37--45 (2015)

\end{thebibliography}
\clearpage
\appendix
\section{Appendix}
\subsection{Modes of Variation}\label{modes_of_variation}
\begin{figure}[!ht]
    \centering
    \includegraphics[scale=0.2]{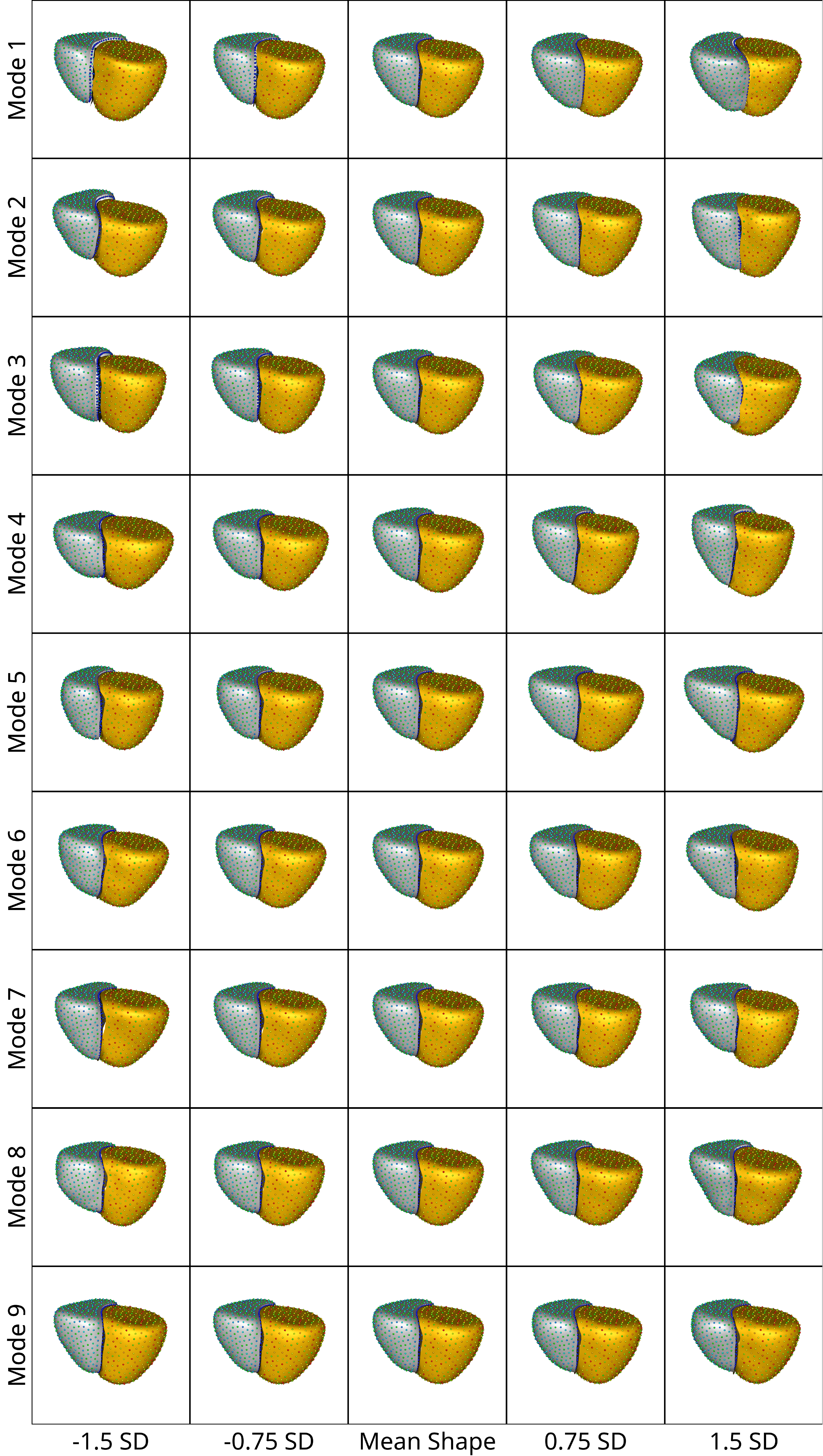}
    \caption{Modes of variation discovered by the shape model of the biventricle dataset with the proposed optimization for multi-organ anatomies with shared boundaries.}
    \label{figure_modes_of_variation}
\end{figure}

\end{document}